\documentclass[12pt]{article}
\usepackage{amsmath}
\usepackage{graphicx,psfrag,epsf}
\usepackage{enumerate}
\usepackage{natbib}
\usepackage{url} 
\usepackage{parskip}
\usepackage{rotating}
\usepackage{multirow}
\usepackage{booktabs}

\newcommand{\blind}{0}

\begin{document}
\def\spacingset#1{\renewcommand{\baselinestretch}%
{#1}\small\normalsize} \spacingset{1}

\if0\blind
{
  \title{\bf Excess deaths and Hurricane María}
  \author{Michael Spagat \and Stijn van Weezel}
    \date{ }
  \maketitle
} \fi

\if1\blind
{
  \bigskip
  \bigskip
  \bigskip
  \begin{center}
    {\LARGE\bf Excess deaths and Hurricane María}
\end{center}
  \medskip
} \fi

\bigskip
\begin{abstract}
We clarify the distinction between direct and indirect effects of disasters such as Hurricane María and use data from the Puerto Rico Vital Statistics System to estimate monthly excess deaths in the immediate aftermath of the hurricane which struck the island in September of 2017.
We use a Bayesian linear regression model fitted to monthly data for 2010--16 to predict monthly death tallies for all months in 2017, finding large deviations of actual numbers above predicted ones in September and October of 2017 but much weaker evidence of excess mortality in November and December of 2017.  
These deviations translate into 910 excess deaths with a 95 percent uncertainty interval of 440 to 1,390. 
We also find little evidence of big pre-hurricane mortality spikes in 2017, suggesting that such large spikes do not just happen randomly and, therefore, the post-hurricane mortality spike can reasonably be attributed to the hurricane.
\end{abstract}
\noindent

\spacingset{1.45} 

\section{Introduction}
There has been controversy over the number of deaths in Puerto Rico caused by Hurricane María, which struck the island in September 2017.
For many months the official death toll was frozen at 64 while independent researchers, working with preliminary data from the Puerto Rico Vital Statistics System (PRVSS), attributed 800 or more excess deaths to the disaster and its aftermath \citep{robles2017official,rivera2018estimating,santos2018use,santoslozada2018estimates}.
Governor Ricardo Rosello responded to this dispute by retaining a team at George Washington University (GWU) that would, he hoped, resolve the large numerical discrepancies.
The GWU effort was pre-empted in May 2018 by \citet{kishore2018mortality} who released a highly publicized survey-based estimate of 4,645 excess deaths with a 95\% uncertainty interval (UI) of 793 to 8498.
This dramatic intervention seems to have prompted the government of  Puerto Rico to release more vital statistic data and to update its own official figure to 1427 excess deaths \citep{puertorico2018transformation}, albeit while providing little information on how they arrived at this number.
\citet{santos2018use,santoslozada2018estimates} used the updated data to estimate 1,139 excess deaths.
This estimate came out remarkably close to the in-depth examination of the GWU team published in November 2018 which, accounting for migration, found 1191 excess deaths (95\% UI: [836; 1544]) between September 2017 and February 2018 \citep{santos2018differential} --- see also \citet{milken2018ascertainment}.
\citet{cruz2019causes} investigate causes of excess deaths within a framework that gives an estimate of aggregate numbers that is consistent with the mainstream, including our own estimates.

In the present paper we estimate excess mortality, fitting a Bayesian linear regression model to the monthly official deaths tallies, from the PRVSS, covering the period 2010--17.
Specifically, we use the 2010--16 data to fit the model and use the estimated parameters to predict the death tallies for each month of 2017, including those before the hurricane, complete with uncertainty intervals (UI).
We then subtract the actual tallies from our predicted tallies. 
For the post-hurricane months we interpret these differences as excess deaths.
Note that in our terminology "prediction errors" and "excess deaths estimates" are interchangeable during the post-hurricane months. 
Of course, there can be no excess deaths prior to the hurricane but we use the differences between predicted and actual values during this period in 2017 to help us gauge how anomalous the post-hurricane differences are.   
The results show a large difference between observed and predicted values for September 2017, with the reported death tally far outside the range predicted by the model. 
For the months September--October we estimate a total of 910 excess deaths with a 95 percent UI of 440 to 1400 deaths (rounding everything to the nearest 10). 
The absence of large differences between observed and predicted values for November and December of 2017 suggests that there was no substantial excess mortality during these two months. 
These results are broadly consistent with previous findings cited above, aside from those of \citet{kishore2018mortality} and the figure of 64, the latter of which seems to have fallen out of circulation. 

We also perform placebo tests, the idea of which is the following. 
We apply our methodology for estimating excess deaths to months before the hurricane actually happened. 
It is impossible for the hurricane to cause excess deaths before it even existed so we hope and expect that the differences between observed and predicted values for the pre-hurricane months will cluster around 0, aside from some random fluctuation. 
This procedure is analogous to testing whether improvements in health that followed patients' consumption of a new drug were truly caused by the drug by determining whether similar improvements followed the consumption of a placebo in a control group. 
We find that no other month between February and August of 2017 displays differences between observed and predicted values at all close in magnitude to those for September and October, although January does display a rather large difference.
These null results for the pre-hurricane period of February through August and, to a lesser extent, for January are reassuring because they suggest that huge prediction errors do not tend to just happen randomly and for no reason.  Thus, they give us confidence that our excess-deaths findings for September and October of 2017 truly reflect reality rather than being mere artefacts of random fluctuations. 

There exists a fairly large and growing literature on excess deaths attributed to wars (e.g. \citet{roberts2004mortality,burnham2006mortality,coghlan2006mortality,degomme2010patterns}, strong claims for the accuracy of such estimates \citep{wise2017epidemiologic} and even proposals to incorporate estimates of excess war deaths into the Sustainable Development Goals (SDG)framework \citep{alda2017beyond}. 
Pre-war death rates have provided the typical counterfactual baseline for these estimates although one prominent estimate used, unconvincingly (see \citet{human2011human}), the death rate for the whole of Sub-Saharan Africa \citep{coghlan2006mortality} as a counterfactual baseline. 
These estimates are plagued by exceptionally wide error bands, sometimes approaching plus or minus 100 percent \citep{roberts2004mortality,burnham2006mortality,coghlan2006mortality}, and a methodological defect that conflates violent deaths caused directly by war with non-violent deaths indirectly attributable to war \citep{spagat2017half,spagat2018terms,spagat2018estimating}. 
High-profile studies have been further discredited due to unrepresentative samples \citep{coghlan2006mortality,human2011human} and fabricated data \citep{burnham2006mortality,spagat2010ethical}.
Yet recent surveys of estimates of excess war deaths take an uncritical \citep{alda2017beyond}, even laudatory \citep{wise2017epidemiologic}, view while ignoring the uncertainty and underplaying the controversy that surrounds these estimates.
Some strong claims in the literature, e.g., that non-violent war deaths dwarf violent ones \citep{wise2017epidemiologic} or that excess-death point estimates should be incorporated into SDG's \citep{alda2017beyond}, are seriously undermined by a realistic appraisal of the uncertainty surrounding excess war-death estimates.

For several reasons, the case of Hurricane María provides a good learning opportunity for the war-deaths literature in addition to the intrinsic importance of understanding the event itself.
First, in contrast to many war settings for which death rates must be estimated through surveys, we have solid, if not perfect, month-by-month data on deaths in Puerto Rico both before and after the hurricane.
Yet we can also observe a quasi experiment in which a methodology typical of that used in the war-deaths literature was implemented in Puerto Rico by \citet{kishore2018mortality} who used a survey to estimate both pre and post hurricane death rates before reliable post hurricane data became available.  The results of \citet{kishore2018mortality} was a central estimate well above what we know now to be a reasonable number and an extremely wide uncertainty interval, the latter of which is a hallmark of the war death literature. 
Second, the counterfactual thinking appropriate for assessing excess deaths in Hurricane María is relatively simple compared to war-based counterfactuals.  This is true, despite the fact that the treatment of migration after the hurricane has a potentially large effect on excess death estimates several months after the event, as \citet{santos2018differential} point out.
Nevertheless, realistic war counterfactuals are likely to also include migration in addition to further complicating factors such as pre-war droughts, coups or economic sanctions. 
Thus, we might expect estimates of excess deaths resulting from Hurricane María to provide an upper bound on the quality attainable for estimates of excess deaths resulting from wars.  

The paper proceeds as follows.
First, we briefly clarify the distinction between deaths caused directly by the hurricane and deaths caused indirectly by the hurricane. 
Next we present our model and estimates. We then compare our methodology with those of some of the other excess deaths estimates.
Finally, we argue for the application of our methods for measuring excess deaths in future violent events.

\section{Excess mortality: A conceptual clarification}
It is important to distinguish between deaths caused directly by the hurricane and deaths caused indirectly by the hurricane.
If the hurricane blew down a tree which then struck a man and killed him that would be a direct death. 
Such deaths are relatively easy to identify and tallying them up is a perfectly valid exercise. 
Yet it is unlikely that direct deaths will capture the full lethal effect of the hurricane. 
It is plausible, for example, that just after the hurricane \emph{X} out of every 100 heart attack victims died while just before the hurricane only \emph{Y} out of 100 died where $X > Y$. 
If so then we can speak of an "excess" of heart attack victims due to the hurricane.
And, of course, we can extend this excess deaths concept to encompass deaths from all immediate causes, not just heart attack deaths. 
When post-hurricane death rates systematically exceed pre-hurricane rates then we can view the difference as hurricane-caused excess.
All the above-quoted studies do this.

We should recognize that our estimates of excess deaths are in the aggregate and not at the individual level. Excess death models, including our own, do not predict whether any particular death is an 'excess' death but rather whether, in the case of Puerto Rico, the death count increased above previous expectation after the hurricane.
Consider, for example, a particular heart attack victim who died before reaching a hospital amidst a shortage of ambulance drivers during the hurricane’s aftermath. 
Would this man have survived if the hurricane had never happened? 
There may be no clear and convincing answer to this question.
This man may have died anyway, even in the best of circumstances.
Or maybe he could have been saved if an ambulance had arrived unusually quickly and contained an unusually experienced paramedic.
There can be many relevant factors governing the fates of particular people and even analysts with good inside information, such as the GWU team, will struggle to identify particular individuals whose fate flipped from survival to death because of, and only because of, the hurricane. 
Thus, a big advantage to the statistical approach to measuring excess deaths is that it operates in terms of probabilities rather than certainties as is appropriate for an environment within which yes-no answers are elusive. In contrast, the intentions of many people in Puerto Rico to create a list of all the people killed by the hurricane (see \citet{sutter2018puerto} and  \citet{santiago2018puerto}) are misguided because there will be a large number of ambiguous cases.

\section{Data}
\begin{figure}[!ht]\centering
  \includegraphics[scale=.5]{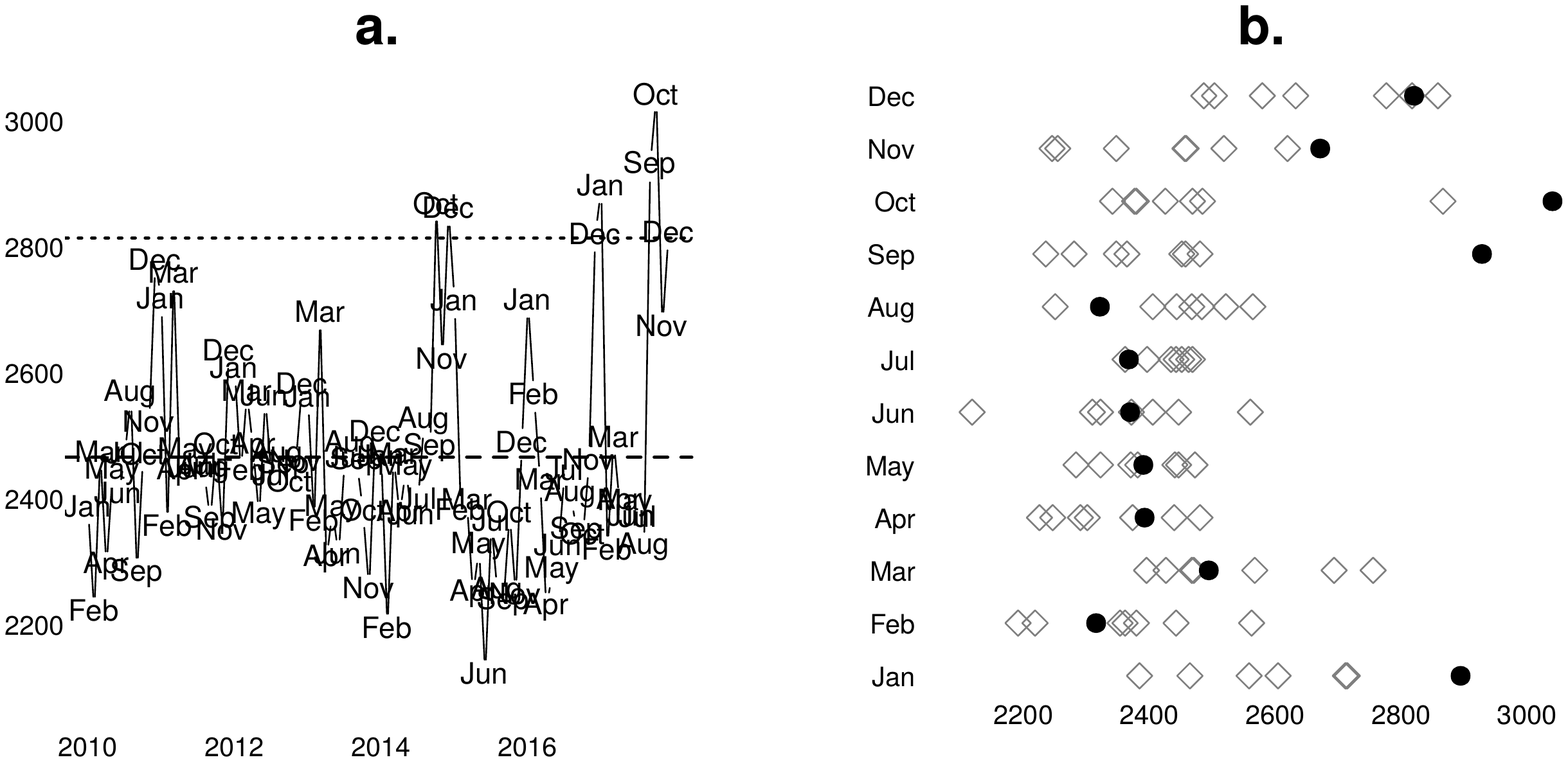}
  \caption{Number of deaths per month, Puerto Rico 2010--17.
  \newline \textbf{a.} the dashed line denotes sample average; the dotted line denotes sample average plus two standard deviations.
  \newline \textbf{b.} solid dots denote data for 2017; diamonds denote data for 2010--16. 
  \newline \emph{Source:} \citet{santoslozada2018estimates,santos2018use,varela2018data}.}
  \label{fig:raw}
\end{figure}

We estimate excess mortality associated with Hurricane María using data for 2010--16 that was originally from the Puerto Rico Vital Statistics System (PRVSS), obtained by Alexis Santos-Lozada of Pennsylvania State University (see \citet{santos2018puerto,santos2018use,santoslozada2018estimates}) and published in Latino USA \citep{varela2018data}.
This information is combined with 2017 data published in \citet{santoslozada2018estimates}.
\citet{santos2018use} acquired the data from the Puerto Rico Institute of Statistics (PRIS) which is a watchdog agency for the official statistics of Puerto Rico, including the PRVSS. 
Ideally the PRVSS would provide these numbers directly, including timely updates. 
However, the PRVSS has stopped releasing data and is not responding to emails according to Santos-Lozada (personal communication).
Meanwhile, the PRIS released the most recent figures in the face of a possibly debilitating reorganization \citep{guglielmi2018plan,wade2018critics}. 
We have no original insight into the behind-the-scenes statistical drama in Puerto Rico which is important but which has no practical effect on any of the excess deaths estimates made using the PRVSS data.
Nevertheless, it is worth noting that the PRVSS numbers used by \citet{santos2018use,santoslozada2018estimates} in their analysis are extremely close, but do not exactly match, the numbers released by PRIS, even for early months.
We did not repeat our analysis using the earlier PRVSS numbers because the PRVSS never released an October 2017 number: \citet{santos2018use,santoslozada2018estimates} impute the October tally in their study.
These data discrepancies must be resolved before there can be a fully definitive account of hurricane-related deaths in Puerto Rico.

\begin{figure}[!ht]\centering
  \includegraphics[scale=.5]{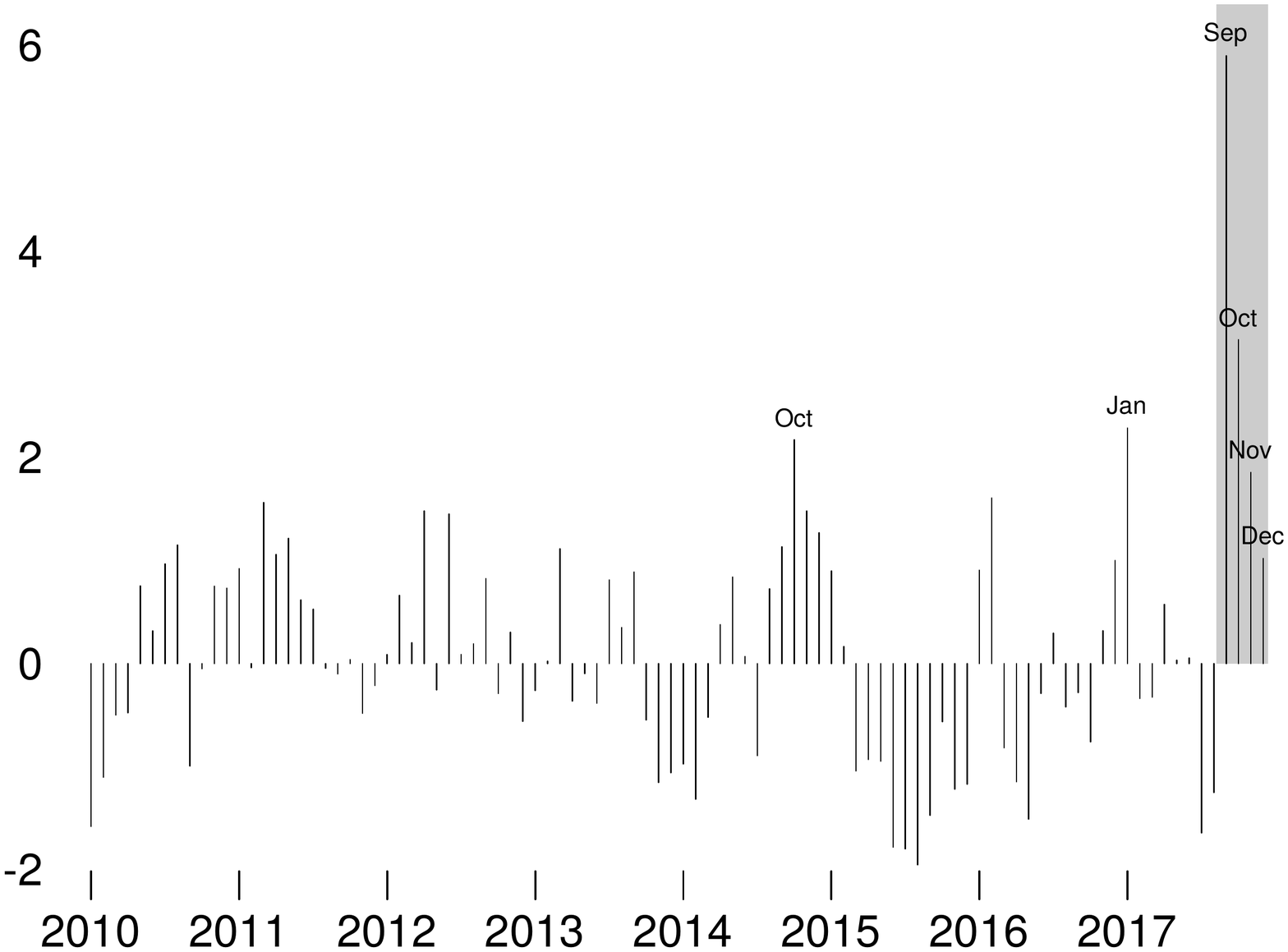}
  \caption{Standardised number of deaths, Puerto Rico 2010--17.   
  \newline \emph{Notes:} Standardised data constructed by subtracting the mean for 2010--16 from the observed number of deaths and dividing by the standard deviation (2010--16). Grey area indicates the months following Hurricane María (Sep-Dec). 
  \emph{Source:} \citet{santoslozada2018estimates,santos2018use,varela2018data}; authors' calculations.}
  \label{fig:standardised}
\end{figure}

Figure~\ref{fig:raw}\emph{a} displays the data, showing the monthly death tallies for the period January 2010 to December 2017. 
There are remarkably high numbers for September and October 2017 compared to historical patterns (fig.~\ref{fig:raw}\emph{b}).
Indeed, the September and October 2017 death tallies are about 550 deaths above the averages for the 2010--16 period (table~\ref{table:statistics}).
The data also shows that the number of deaths in January 2017 --- well before the hurricane struck the island --- is also significantly above average. 

To gain better insight into the potential impact of the hurricane on mortality we standardize the time-series data to account for historical variation. 
Specifically, for each observation $x_t$ we subtract that month's mean $\mu_m$ during the baseline period (2010--16) and divide by that month's standard deviation to obtain $\sigma_m$: $z_{t} = (x_{t}- \mu_m) / \sigma_m$.
The standardised data show September and October 2017 as large positive anomalies (Fig.~\ref{fig:standardised}), although there are a few other large anomalies in the time-series data, most notably October 2014 and the before-mentioned January 2017. 
Additionally, there are some curious runs of positive and negative anomalies for which there is currently no explanation and that deserve attention in subsequent work. 
They are, however, beyond the scope of this study.

\begin{table}[!ht]\centering
\caption{Reported deaths Puerto Rico 2010--17.}
   \label{table:statistics}
    \begin{tabular}{lccc}    
    ~            & Average 2010--16 & 2017 & Difference\\
    \\[-1.8ex]\hline \\[-1.8ex]
    January      & 2,592 (132) & 2,894     & 302\\
    February     & 2,358 (127) & 2,315     & -43\\[-1.8ex]\\
    March        & 2,539 (138) & 2,494     & -45\\
    April        & 2,336 (97)  & 2,392     & 56\\
    May          & 2,388 (69)  & 2,390     & 2\\[-1.8ex]\\
    June         & 2,362 (137) & 2,369     & 7\\
    July         & 2,431 (39)  & 2,367     & -64\\
    August       & 2,448 (101) & 2,321     & -127\\[-1.8ex]\\
    September    & 2,373 (94)  & 2,928     & 555\\
    October      & 2,477 (179) & 3,040     & 563\\
    November     & 2,414 (139) & 2,671     & 257\\
    December     & 2,664 (152) & 2,820     & 156\\
    \\[-1.8ex]\hline \\[-1.8ex]    
    \multicolumn{4}{p{20em}}{\emph{Notes:} Standard deviation in parentheses.
    \newline \emph{Source:} \citet{santoslozada2018estimates,santos2018use,varela2018data}}
    \end{tabular}
\end{table}

\section{Empirical framework}
Bayesian methods are applied to analyse the data and estimate the number of excess deaths. 
The model can be denoted in terms of a normal likelihood function as: 

\begin{align}
y_{t} \sim \mathcal{N}(\alpha + \theta_m x_{t}, \sigma^2_y), \; \text{for}\ t=1,...,N  
\end{align}

where $y_t$ is the actual reported number of deaths at time $t$ and $x_t$ indicates the month of the year.
$\alpha$ is a constant which, in this case, represents the estimated number of deaths in the baseline month (January) while $\theta_m$ is the estimated difference in deaths relative to the baseline (i.e. January) for each individual month (Feb--Dec). 
In other words, we regress the number of deaths ($y_1,...,y_n$) on a vector of month indicators ($x_1,...,x_n$) omitting the baseline month so that the model is not overdetermined. 
Due to the time-series nature of the data the errors ($\sigma^2_y$) are adjusted for serial correlation using an $AR(1)$ correction; modeling the error as $\epsilon_t = \rho_{t-1} + \eta_t$. 

The model's key feature is that the predicted (or fitted) number of deaths $y_i$ depends on the month of the year through the estimated parameter distribution $\theta_m$.
As such it accounts for systematic differences between months due to, for instance, seasonality. 
The model is fitted to the data covering the years 2010--16 and the estimated posterior distribution is subsequently used to generate predictions for the expected number of deaths for each month in 2017, including the pre-hurricane months which serve to check the predictions against the data.  
The predicted values for September--December serve as a counterfactual for a hypothetical world in which the hurricane never happened. 
For the pre-hurricane months (January--August) the predicted values serve as a placebo test; the expectation is that the predictions are close to the historical ranges subject to some random variation. 
In short, the excess mortality estimates are based on deviations in post-hurricane monthly tallies from the observed historical patterns for each month. 

One constraint the analysis faces is that there is, unfortunately, only a limited amount of information to construct a baseline. 
The available data only goes back to 2010 and, in any case, the further we go back in time the less useful the data potentially are for predicting the future. 
Given the relatively small sample size ($N=84$) it is important to account for uncertainty. 
An advantage of the Bayesian framework is that it efficiently incorporates all the available information, including in small samples \citep{zellner1988optimal}. 
Importantly, this framework treats the data as fixed and the parameters random. 
Thus, under the sampling-based Bayesian method used in this study (Hamiltonian Monte Carlo), the quality of the inference is controlled not by asymptotic properties but by the number of samples taken \citep{kruschke2014doing}.  

Ideally in a small sample an informative prior is specified as an initial estimate \citep{dunson2001commentary,kadane2015bayesian,mcneish2016using}.
However, given that all the available data is needed to estimate the parameters, this means that no information is left to formulate informative priors \citep{gelman2017prior}. 
Therefore, we use a diffuse prior distribution that relies on the default settings in R's 'brms' package \citep{burkner2017brms} --- an interface for Stan \citep{carpenter2017stan} --- which follow a Student $t$ distribution ($t(3,0,1)$).
This entails that the estimates will be close to what would be obtained using frequentist methods and, additionally, that they are sensitive to the idiosyncrasies of the data. 
The parameters are estimated using 4 Markov chains with 2,000 iterations each, the first 1,000 of which serve as warm-up. 
The model converged based on visual inspection of the trace plots and the $\hat{R}$ statistics of 1. 

\section{Empirical results}
\begin{figure}[!ht]\centering
  \includegraphics[scale=.5]{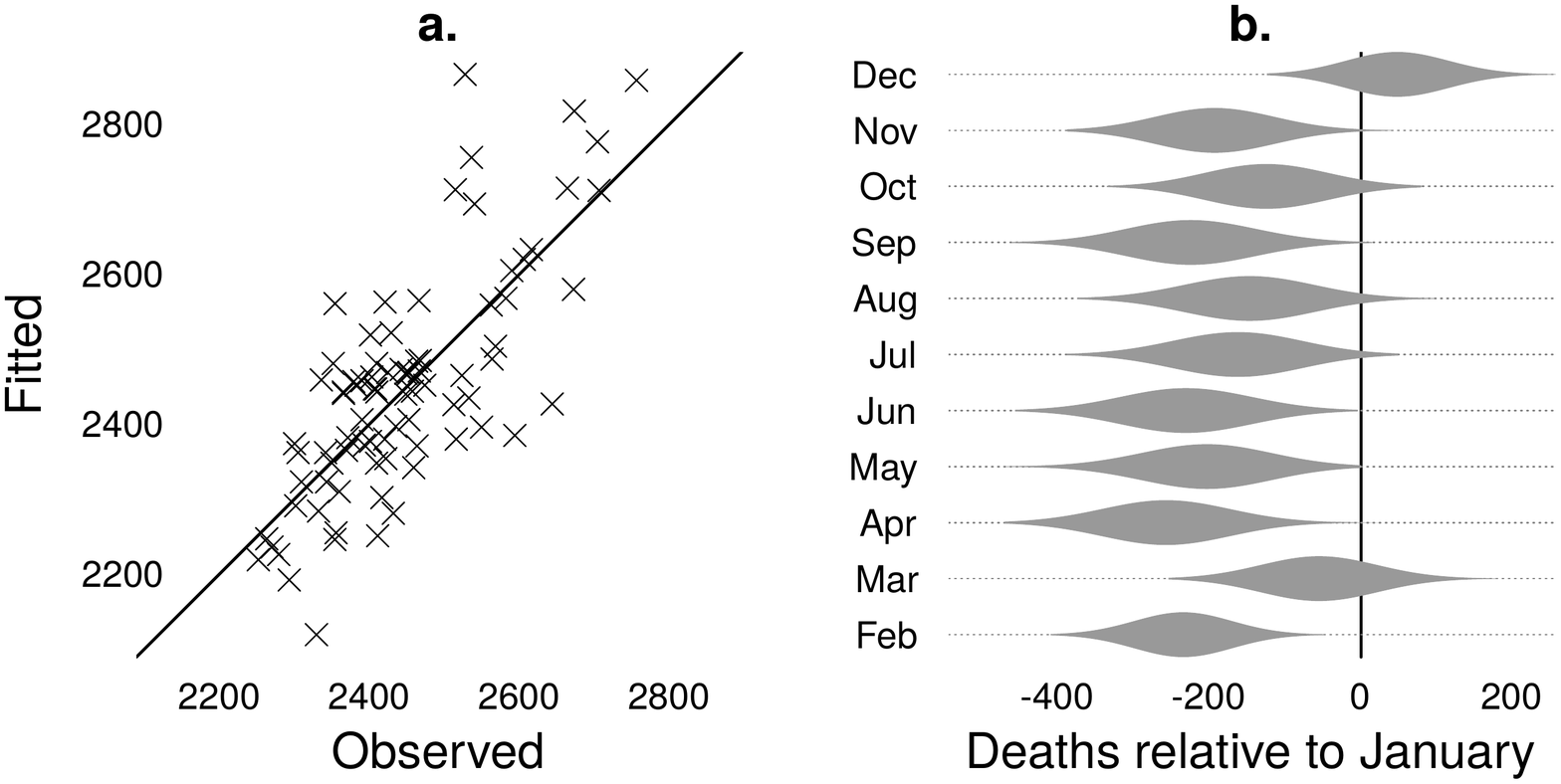}
  \caption{Regression results.
  \newline \textbf{a.} fitted data points are means of the posterior distribution; solid line is the 45-degree line.
  \newline \textbf{b.} displays parameters' posterior distribution, excluding the intercept (January).}
  \label{fig:posterior}
\end{figure}

Figure~\ref{fig:posterior}\emph{a} displays the relatively good fit of the model, comparing the observed values with the fitted values --- using the mean of the posterior distribution as a central estimate. 
The model's root mean squared error (RMSE) of 98 is considerably better than the RMSE of 114 obtained without adjusting errors for serial correlation (table~\ref{table:results}, model 2.) or the RMSE of 130 obtained in a dynamic model linking deaths to the lagged outcome variable and seasonal indicators (table~\ref{table:results}, model 4.) . 
For reference, a parsimonious intercept-only model has an RMSE of 150. 
The model also performs better when considering the Leave-One-Out Information Criterion as a measure for goodness-of-fit \citep{vehtari2017practical}. 
Figure\ref{fig:posterior}\emph{b} displays the estimated posterior distribution and illustrates that the number of deaths is relatively low during the hurricane season (June--November) compared to the January baseline.
Mortality tends to be higher during the dry season (December--March).  

\begin{figure}[!ht]\centering
  \includegraphics[scale=.5]{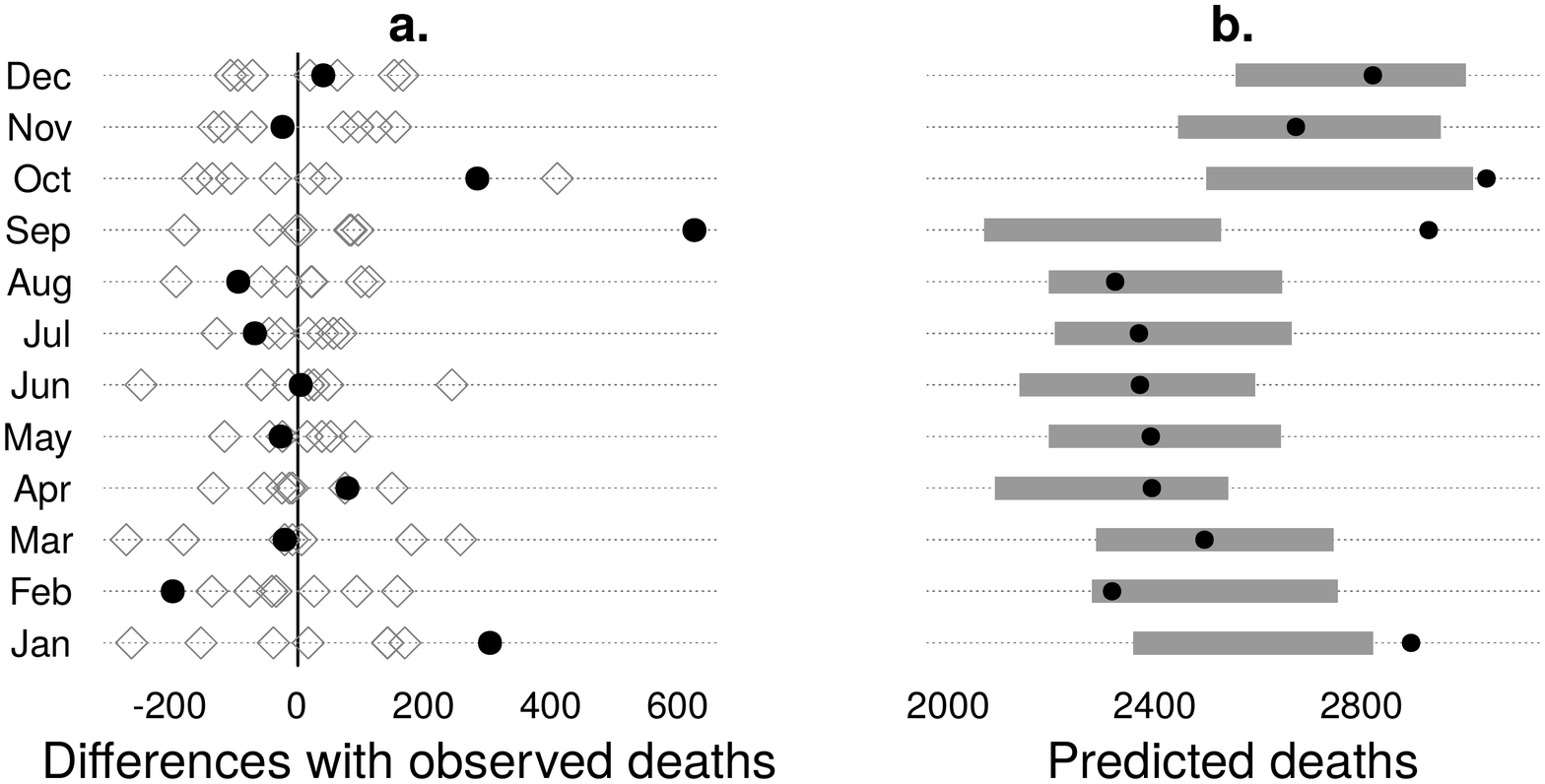}
  \caption{Predicted/fitted deaths and excess mortality.
  \newline \textbf{a.} displays differences between observed numbers of deaths and the means of the predicted posterior distribution; diamonds denote data for 2010--16; solid dots denote data for 2017. 
  \newline \textbf{b.} displays 95\% UI with observed data for 2017 denoted by solid dots.}
  \label{fig:estimates}
\end{figure}

The parameters' posterior distribution is used to predict the outcome (number of deaths) for 2017.  
Figure~\ref{fig:estimates}\emph{a} displays the difference between the observed number of deaths and the mean of the predicted posterior distribution for each respective month in 2017 (closed dots) and other years (open dots).  
The estimates for the years 2010--16 are obtained in similar manner to those for 2017: one year is left out at a time and the outcome is predicted using the parameter estimates based on the data from the included years. 

\begin{table}[!ht]\centering
\caption{Fitting number of deaths at time $t$ --- Linear regression.}
   \label{table:results}
   \scalebox{.8}{
    \begin{tabular}{lcccc}
    Variable         & Model 1        & Model 2   & Model 3   & Model 4 \\
    \\[-1.8ex]\hline \\[-1.8ex]
    Intercept        & 2449 (17)      & 2591 (46) & 2593 (47) & 1813 (260)     \\[-1.8ex]\\
    February         & ~              & -233 (65) & -236 (48) & ~              \\
    March            & ~              & -51 (65)  & -55 (57)  & ~              \\
    April            & ~              & -254 (66) & -257 (61) & ~              \\[-1.8ex]\\
    May              & ~              & -202 (67) & -205 (64) & ~              \\
    June             & ~              & -229 (66) & -231 (65) & ~              \\
    July             & ~              & -160 (65) & -162 (64) & ~              \\
    August           & ~              & -143 (66) & -146 (64) & ~              \\[-1.8ex]\\
    September        & ~              & -217 (66) & -223 (63) & ~              \\
    October          & ~              & -113 (65) & -122 (62) & ~              \\
    November         & ~              & -177 (65) & -191 (57) & ~              \\
    December         & ~              & 74 (66)   & 50 (48)   & ~              \\[-1.8ex]\\
    Lagged outcome   & ~              & ~         & ~         & 0.2 (0.1)      \\
    Hurricane season & ~              & ~         & ~         & 62 (42)        \\
    Dry season       & ~              & ~         & ~         & 169 (45)       \\[-1.8ex]\\
    $LOO-IC$         & 1085 (15)      & 1062 (15) & 1040 (17) & 1055 (17)      \\
    $RMSE$           & 150            & 114       & 98        & 130            \\[-1.8ex]\\
    $N$              & 84             & 84        & 84        & 83             \\
    Adjusted $\sigma^2$ & N           & N         & Y         & ---            \\
    \\[-1.8ex]\hline \\[-1.8ex]
    \multicolumn{5}{p{30em}}{\emph{Notes:} Table presents the mean of the posterior distribution (standard deviation in parentheses).
    \newline Estimates are based on 4 chains with 2,000 iterations each, first 1,000 serve as warm-up. 
    Model fitted using R's 'brms' package \citep{burkner2017brms} applying a diffuse prior: $t(3,0,1)$.
    \newline 'Hurricane season' is a binary indicator taking value 1 for the months June--November and 0 otherwise; 'Dry season' is a binary indicator taking a value 1 for the  months December--March and 0 otherwise.
    \newline $LOO-IC$, leave-one-out information criterion; $RMSE$, Root mean squared error; $N$, number of data points.
    \newline $\sigma^2$ adjustment is for $AR(1)$ autocorrelation as described in text.}    
    \end{tabular}
    }
\end{table}
\clearpage

We make the following observations. 
First, the average predicted value for September 2017 shows a huge difference, suggesting that there were many hurricane-related deaths. 
There is a large discrepancy for October 2017 as well, although the data shows an earlier large and positive October value in 2014. 
This introduces some uncertainty as to whether the full difference for October 2017 can be attributed exclusively to the hurricane. 
Second, the predicted values for November and December 2017 fall within the range of expected deaths based on historical patterns. 
As such, there is little evidence pointing towards substantial hurricane-related deaths during the last two months of 2017. 
Third, and finally, the remaining 2017 observations look routine except for a rather large positive value for January which, naturally, cannot be attributed to the hurricane. 

To construct the main excess deaths estimates the predicted distribution is subtracted from the reported mortality number for September and October 2017. 
Table~\ref{table:estimates} presents these results, reporting the central estimate, i.e., the mean of the distribution, along with the 50\% and 95\% uncertainty intervals, the former of which are computationally more stable than the latter (predictions based on parameters from model 3 in table~\ref{table:results}).
All the numbers are rounded to the nearest ten.  
The estimates show a central estimate of 910 excess deaths for September and October combined with a 95\%UI from 430 to 1,400.
This combined two-month estimate breaks down into a 620 central estimate for September (95\%UI:[400; 850]) and 280 for October (95\%UI:[40; 540]).
These estimates do change a bit when we treat the October 2014 value as an outlier and exclude it from the data. 
The estimated excess mortality in this case increases slightly to 970 (95\% UI: [510;1430]); the October estimate increases to 340 (95\%UI:[90;580]).

As an aside, as previously explained, the estimates are based on using a diffuse prior given the limited amount of data.
However, Bayesian estimation should provide accurate estimates in small samples with a proper informative prior \citep{kadane2015bayesian}; although we do have to accept the fact that the prior could influence the results of the analysis \citep{mcelreath2018statistical}.
Casting aside our reservations, we therefore conduct a sensitivity check specifying an informative prior based on observed mortality data for 2010--11, estimate the parameters using data for 2012--16, and subsequently predict the outcome for 2017. 
The results are qualitatively and quantitatively similar.
The estimated number of excess deaths for September and October combined is 880 with a 95\% UI of 390 to 1390.
This breaks down into a central estimate of 620 deaths for September (95\%UI:[390; 870]) and 260 for October (95\%UI:[0; 530]).

\begin{table}[!ht]\centering
\caption{Estimated excess deaths}
   \label{table:estimates}
    \begin{tabular}{lcccc}    
    ~                & September & October  & Sep--Oct          & Sep--Dec \\
    \\[-1.8ex]\hline \\[-1.8ex]
    Central estimate & 620       & 280      & 910               & 920                \\
    50\% UI          & 550; 700  & 200;370 & 740; 1070         & 600; 1250          \\
    95\% UI          & 400; 850  & 40; 540  & 440; 1400         & -30; 1900          \\
    \\[-1.8ex]\hline \\[-1.8ex] 
    \multicolumn{5}{p{30em}}{\emph{Notes:} Predictions based on parameters from model 3 in table~\ref{table:results}.}
    \end{tabular}
\end{table}

To test the strength of the estimates a placebo test is conducted on the pre-hurricane months (January through August) in 2017. 
The essential idea is that there was no major event (treatment) during these months so we would expect that the death tallies for January--August 2017 to fluctuate within the predicted ranges during this period. 
If this assumption is violated --- i.e. if we observe some large predictions errors during the pre-hurricane period ---, then we should discount our belief in the claim that the post-hurricane prediction errors signify excess deaths attributable to the hurricane. 
In other words, if huge random fluctuation in the monthly death tallies are routine then it would seem plausible that the post-hurricane prediction errors are just part of this routine of substantial fluctuations rather than being caused by Hurricane María. 
However, the results show that the 2017 monthly tallies for February through August all fall well within the 95\% UI of the predicted posterior distribution as displayed by figure~\ref{fig:estimates}\emph{b}. 
One exception to this is the January tally which fails this placebo test by exceeding the upper bound of its 95\% UI by about 70 deaths. 
This is of course just one test out of eight and not a resounding failure; it could be that the January 2017 value is a random occurrence. 
Perhaps something did occur during January 2017 which nudged the death tally for that month beyond the range of predicted deaths based on historical mortality patterns. 
Pending further investigation this estimate should be interpreted as a small warning sign that the excess death estimate could potentially be on the high side. 

Focusing on a larger period, covering September through December 2017, leads to an excess deaths estimate of 920 (95\%UI:[-40; 1870]). 
This central estimate is close to the September-October value but the bottom of the UI now dips below 0.  
As already noted, the November and December prediction errors look to be within the range of normal fluctuations. 
Applying the second type of placebo test shows that the average prediction error for September-December 2017 is 2.3 times larger than the average prediction error for September-December 2010--16.
This figure is substantial larger than 1 but still well below the figure of 4.5 obtained for the September--October estimate. 
Generally, the estimates for November and December provide little empirical evidence for excess mortality attributable to Hurricane María.

\begin{figure}[!ht]\centering
  \includegraphics[scale=.5]{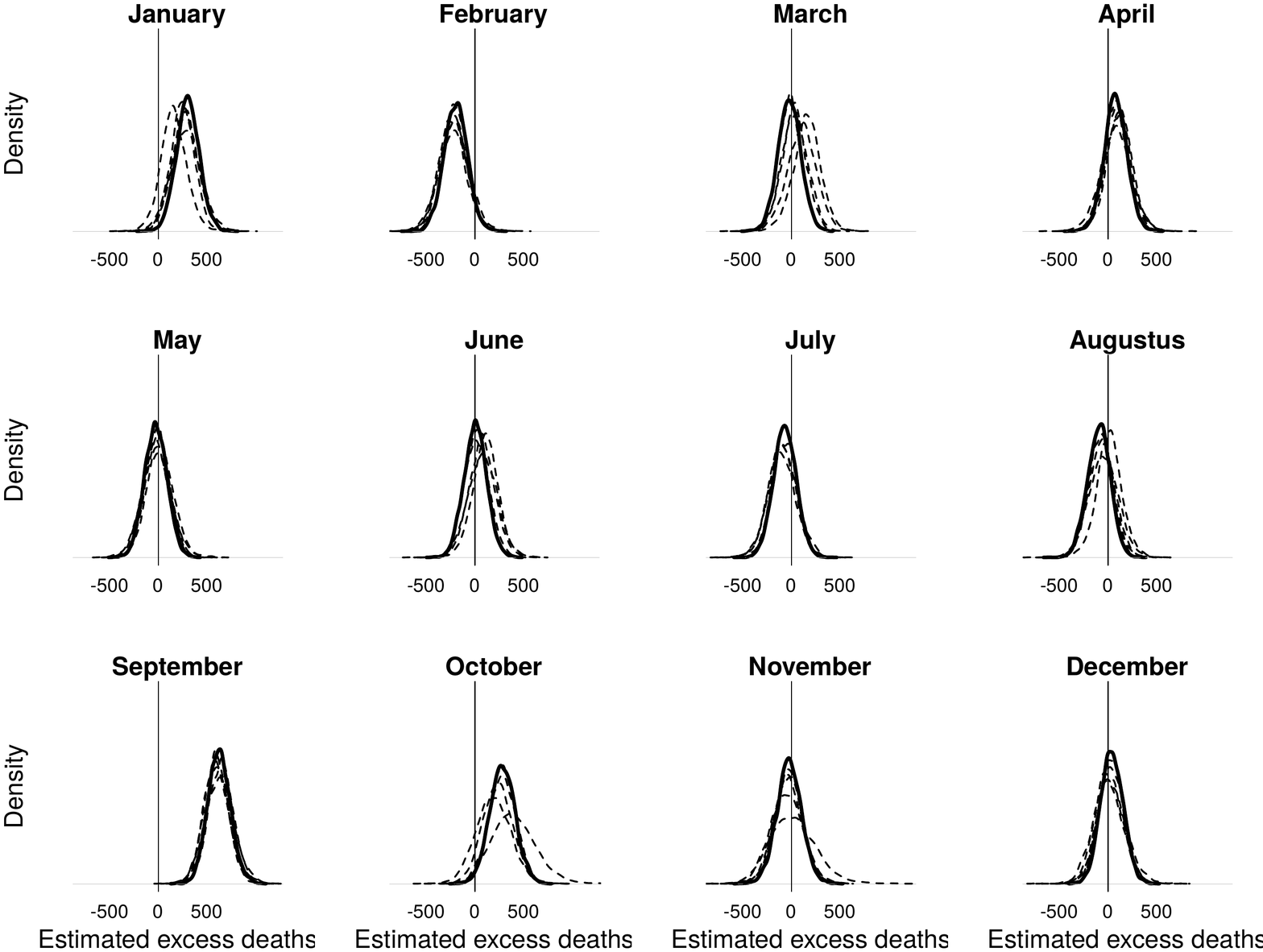}
  \caption{Distribution of estimated excess deaths --- Different baselines.
  \newline \emph{Notes:} Solid line indicates the main estimate based on parameters fitted to data for 2010--16; dashed lines indicate estimates base on parameter using different baseline periods excluding years between 2010 and 2015 one at a time. For the post-hurricane months (Sep--Dec) the difference between the predicted and observed number of deaths is interpreted as an estimate of excess mortality.}
  \label{fig:benchmark}
\end{figure}

Additional placebo tests are carried out by re-estimating the model and generating predictions for the number of deaths in September--October for each year between 2010 and 2016, leaving out one year at a time. 
More specifically, the model is fitted covering the years 2010--15 and predictions generated for 2016; subsequently the model is fitted covering the years 2010--14, 2016 and predictions generated for 2015, etc. 
This exercise shows that the average prediction error for September--October 2017 is about 4.5 times larger than the average prediction error for September--October 2010--15: 450 versus 100. 
Only considering one-year ahead predictions --- e.g. predicting the outcome for 2015 based on 2010--14 data excluding 2016, as a reviewer kindly suggested, produces similar estimates; the prediction error being 3.8 times larger for September--October 2017 compared to the average prediction error for the other estimates.  
These results are consistent with the idea that the hurricane did indeed cause many additional deaths. 
In addition, the estimates are not sensitive to changing the sample used to construct the baseline or to considering short-term trends in mortality, as illustrated by figure~\ref{fig:benchmark}. 

\section{Comparison with other estimates}
\begin{table}[!ht]\centering
\caption{Summary existing estimates excess deaths}
   \label{table:summary}
    \begin{tabular}{lcc}    
    Study                                           & Estimate & 95\% UI     \\
    \\[-1.8ex]\hline \\[-1.8ex]
    \citet{santoslozada2018estimates,santos2018use} & 1,110    & ---         \\
    \citet{robles2017official}                      & 1,052    & ---         \\
    \citet{rivera2018estimating}                    & 822      & 605-1039    \\
    \citet{santos2018differential}                  & 1,271    & 1,872-2,315 \\
    \citet{kishore2018mortality}                    & 4,645    & 793-8,498   \\
    \\[-1.8ex]\hline \\[-1.8ex]    
    \multicolumn{3}{p{25em}}{\emph{Notes:} See referenced work for details on estimation.}
    \end{tabular}
\end{table}

There are several other studies that have tried to estimate excess mortality in Puerto Rico stemming from Hurricane María (table~\ref{table:summary} provides a summary; see also \citet{sandberg2019all})
\citet{santos2018use,santoslozada2018estimates} are most similar to our work but there are some important differences.
First, they define excess deaths as deviations from the upper limits of their monthly uncertainty interval for predicted death tallies. 
In contrast, we subtract the distribution of predicted death tallies rather than only the upper limits. 
This means that their excess death estimates are what we would view as the lower bound of the uncertainty interval. 
Second, they do not offer uncertainty intervals on the excess deaths estimates as we do. 
Third, their predictions for each month are based only on 2010-2016 data for that month whereas we use all the data in each of our predictions. 
Fourth, we perform placebo tests on the pre-hurricane 2017 months to assess the chances that large positive prediction errors can just happen randomly. 
Fifth, we perform placebo tests on September--October 2010-2016 in search of evidence that these two months may simply be months that are prone to large prediction error.
Thus, there are several important methodological differences between \citet{santos2018differential} and our work. 
Nonetheless, perhaps surprisingly, our central excess death estimate winds up being reasonably close to theirs: 910 versus 1,110.

\citet{robles2017official} use daily data from 42 days following the hurricane --- i.e. from 20 September 2017 through 31 October --- while noting that their October data are likely to be incomplete. 
For each day they subtract the 2015--16 average for that same day from the 2017 number.
Although their excess death concept is similar to ours, they apply it at a daily rather than at a monthly level.
In contrast, we average over more years and our regression approach incorporates data from all months for our predictions.  
We also construct uncertainty intervals and perform placebo tests to build confidence that the September--October prediction errors really are hurricane related.
Again, despite the differences, the \citet{robles2017official} estimate of 1,052 excess deaths is close to ours, although presumably their estimate would be higher if they had more complete October data. 

\citet{rivera2018estimating} is similar to the work by \citet{robles2017official} in focusing on the 42 days following the landfall of María. 
However, \citet{rivera2018estimating} depart from previous work by using a binomial model to estimate the average probability of death following the hurricane, using the 19 days before landfall as a baseline. 
Their maximum likelihood estimate shows a death toll of 822 (95\% UI:[605; 1039]), somewhat lower than the other estimates but their UI overlaps substantially with ours which ranges from 440 to 1390.

Finally, similar to our work \citet{santos2018differential} also use a regression model; fitting data from July 2010 to August 2017 and using the predictions from the fitted model as counterfactual estimates. 
They estimate 1271 excess deaths (95\% UI: [1154; 1383]) for September--October 2017 and 2098 (95\% UI: [1872; 2315]) for September/December 2017. 
Their main unique feature is to factor in displacement and socio-economic determinants of mortality.  These adjustments have only a limited effect on the September--October estimates but quite a big one on estimates for subsequent months.  We believe that future debate should focus on November and beyond where there is still room to discuss the migration-driven counterfactual used by \citet{santos2018differential} which goes beyond the scope of our paper.

Regardless of methodology, these four different approaches do arrive at strikingly similar results for September--October 2017 with estimates in the range of 800 to 1,200 deaths. 
At odds with these estimates is the recent survey of \citet{kishore2018mortality} with a central estimate of 4,645 excess deaths.
Due to the small sample size this estimate comes with a very wide uncertainty interval of 793 to 8,498 deaths, which at least overlaps with the work based on data from the Vital Statistics System, albeit at the cost of rendering this survey work almost meaningless. 
Small samples such as in \citet{kishore2018mortality} ($N=3,299$) leads to erratic estimates that should be regarded cautiously.   

\section{Conclusions}
At this stage the scale of the death toll attributable to Hurricane María during September and October of 2017 is fairly clear. 
There is a broad consensus among multiple studies based on data from the Puerto Rico Vital Statistics System that the excess death toll from the hurricane is probably somewhere around 1,000, with most studies, although not our own, putting the estimate somewhat above this figure. 
Even the initially resistant government of Puerto Rico has broadly recognized the reality discussed above with their updated figure of 1,427, which, surprisingly is on the high end of the independent vital-statistics-based work. 
Unfortunately, the government has not presented details that would allow us to assess this figure although these details may be forthcoming when the GWU team releases further findings.

The only dissonant note comes from the survey-based \citet{kishore2018mortality} study which we think is best viewed as a quasi experiment that compares the survey approach for measuring excess, commonly used in war environments, to a vital-statistics-based approach in an environment where these statistics are highly reliable if not perfect. 
The outcome is sobering for the excess war-deaths literature: a central estimate that is far too high and an very wide uncertainty interval reaching beyond plus or minus 83 percent.  
Also discouraging are the t shirts, hats and placards displaying the number 4,645 at demonstrations in Puerto Rico which can easily be found on the internet.

We caution against projects attempting to make a list of all the particular individuals whose deaths, including non-violent ones, are attributable to the hurricane.  
Causes of death tend to be multiple and varied so while the hurricane will have played a role in many deaths it will often not be the single unique cause of death.  
In other words, the excess death concept is statistically, not individually, based.

We hope that our paper will have a lasting effect on methodologies used to estimate excess deaths following violent events in the future. 
We emphasize the utility of two main points from our methodology. 
First, there is the Bayesian regression approach that efficiently integrates all relevant data. 
Second, there are the placebo tests that help us distinguish between random fluctuations in death tallies and systematic movements in these tallies caused by a violent event such as a hurricane. 
Hopefully, the use of these techniques can reduce the scale of future controversies.  

\bibliographystyle{chicago}
\let\oldbibliography\thebibliography
\renewcommand{\thebibliography}[1]{\oldbibliography{#1}
\setlength{\itemsep}{1pt}} 
\bibliography{references}

\end{document}